\documentclass[A4,aps,twocolumn]{revtex4}
 \usepackage{graphicx}
 \usepackage{color}
 \usepackage[sort&compress]{natbib}
 \usepackage{mathrsfs} 
 \usepackage{amsmath,lscape,epsfig}

  \def\beq{\begin{equation}}
  \def\eeq{\end{equation}}
  \def\beqa{\begin{eqnarray}}
  \def\eeqa{\end{eqnarray}}
  \def\ban{\begin{eqnarray*}}
  \def\ean{\end{eqnarray*}}
  \def\bi{\begin{itemize}}
  \def\ei{\end{itemize}}

\begin{document}

\title{Quark matter nucleation with a microscopic hadronic equation of state} 
  \author{Domenico Logoteta, Constan\c{c}a Provid\^encia, Isaac Vida\~na}
  \affiliation{Centro de F\'{\i}sica Computacional, Department of Physics,
   University of Coimbra, PT-3004-516 Coimbra, Portugal} 
 \author{Ignazio Bombaci}
 \affiliation{Dipartimento di Fisica ``Enrico Fermi'', 
 Universit\'a di Pisa,
 and INFN Sezione di Pisa, 
  Largo Bruno Pontecorvo 3, I-56127 Pisa, Italy }

 \begin{abstract}
 
 The nucleation process of quark matter in cold ($T = 0$) stellar matter  is
 investigated using the microscopic Brueckner--Hartree--Fock approach to describe 
 the hadronic phase, and the MIT bag model, the Nambu--Jona--Lasinio, and 
 the Chromo Dielectric models  to describe the deconfined phase of quark matter. 
 The consequences of the nucleation process for neutron star physics are
 outlined. Hyperonic stars are metastable only  
 for some of the quark matter equations of 
 state considered. The effect of an hyperonic
 three body force on the metastability of compact stars is estimated, and it
 is shown that, except for the  Nambu--Jona--Lasinio model and the MIT bag model
 with a large bag pressure, the other models predict the formation of hybrid stars 
 with a maximum mass not larger than $\sim 1.62 \, M_\odot$.
 \end{abstract}
 
  \maketitle
  
  \vspace{0.50cm}
  PACS number(s): {97.60.-s, 97.60.Jd, 26.60.Dd, 26.60.Kp} 
  \vspace{0.50cm}


\section{Introduction}

Quark matter (QM) nucleation in neutron stars has been studied by many authors 
both at zero \cite{b0,b1,b2,b3,b4,b5,b6,b7,b8} and finite temperature 
\cite{h1,h2,h3,h4,me,me1}, due to its potential connection with explosive
astrophysical events such as supernovae and gamma ray bursts. In all these works, 
the hadronic phase was described using phenomenological models, such as 
{\em e.g.,} the well-known relativistic mean field (RMF) model based on effective 
Lagrangian densities where the baryon-baryon interaction is described in terms of 
meson exchanges \cite{serot86}. Among the different RMF models, one of the most
popular parametrizations is the one of Glendenning and Moszkowski \cite{gm91} of the 
non-linear Walecka model which have been widely used to study the effect of the 
hadronic equation of state (EoS) on the QM nucleation process. In particular, the effect 
of different hyperon couplings on the 
critical mass \cite{b0,b1} for pure hadronic stars (HSs, {\it i.e.} neutron stars 
in which no fraction of QM is present)  
and the stellar conversion energy \cite{grb}
was studied in Ref.\ \cite{b6}. It was found that increasing the value of the hyperon 
coupling constants, increases the stellar metastability threshold mass and the value of the 
critical mass, thus making the formation of 
quark stars (QSs, {\it i.e.} hybrid stars or strange stars depending on the 
details of the EoS for quark matter used to model the phase transition)   
less likely. 
In that work, the hadronic phase was also described using the quark-meson-coupling 
model \cite{qmc}, concluding, in that case, that the formation of a quark star was only 
possible with using a small value of the bag pressure. In all these works the MIT bag model \cite{mit} was 
used to describe the quark matter phase. In a recent work \cite{b9}, two models that 
contain explicitly the chiral symmetry were applied to describe the quark phase, namely 
the Nambu--Jona-Lasinio (NJL) model \cite{nambu} (see also \cite{bba2,njl1}) and the 
Chromo Dielectric model  (CDM) \cite{cdm,cdm1}. It was shown there that it is very difficult 
to populate the quark star branch with the NJL model and, therefore, 
all compact stars would be pure hadronic stars in that case. On the contrary, 
with the CDM, both hadronic and quark star configurations can be formed.

In the present work we study the nucleation of quark matter using an hadronic EoS based on microscopic calculations. 
In particular, we employ two hadronic EoS based on microscopic 
Brueckner--Hartree--Fock (BHF) calculations of hypernuclear matter. The first one (hereafter 
called Model $1$) is the recent parametrization provided by Schulze and Rijken \cite{hans} which 
uses the Argonne V18 nucleon-nucleon (NN) \cite{Ar18} supplemented by the microscopic 
three-body force (TBF) of Ref.\ \cite{hans_NNN}  
between nucleons (NNN), 
and the recent Nijmegen extended soft-core ESC08b hyperon-nucleon (YN) potentials \cite{nij08}. 
The second one (hereafter called Model $2$) is based on our recent work of 
Ref.\ \cite{isaac2011} where we used 
the Argonne V18 NN force and the Nijmegen soft-core NSC89 YN one \cite{ny} in a microscopic 
BHF calculation of hyperonic matter supplemented with additional simple phenomenological 
density-dependent contact terms, that mimic the effect of NNN, NNY and NYY TBFs, 
to establish numerical lower and upper limits to the effect of hyperonic TBF on the 
maximum mass of neutron stars. To describe the quark phase, 
in the present work, we use the three different models already mentioned, the MIT bag 
model \cite{mit}, the NJL model \cite{nambu} and the CDM model \cite{cdm}. 

The paper is organized in the following way. In sections\ \ref{sec:eos} and \ref{sec:pe}, we 
briefly review the BHF approach and the main features of quark matter nucleation in hadronic 
stars, respectively. Our results are presented in Sec. \ \ref{sec:results}. Finally, a summary and 
the main conclusions of this work are given in Sec.\ \ref{sec:conclusions}.

 
\section{The BHF approach}
\label{sec:eos}

The BHF approach is the lowest order of the Brueckner--Bethe--Goldston (BBG) many-body theory \cite{bbg}.   
In this theory, the ground state energy of nuclear matter is evaluated in terms of the so-called
hole-line expansion, where the perturbative diagrams are grouped according to the number of independent
hole-lines. The expansion is derived by means of the in-medium two-body scattering $G$ matrix. The
$G$ matrix, which takes into account the effect of the Pauli principle on the scattered particles
and the in-medium potential felt by each nucleon, has a regular behavior even for short-range repulsions,
and it describes the effective interaction between two nucleons in the presence of a surrounding medium.
In the BHF approach, the energy is given by the sum of only {\it two-hole-line} diagrams including
the effect of two-body correlations through the $G$ matrix. It has been shown by Song {\it et al.}
\cite{song98} that the contribution to the energy from {\it three-hole-line} diagrams (which account
for the effect of three-body correlations) is minimized when the so-called continuous prescription
\cite{jeneuke76} is adopted for the in-medium potential, which is a strong indication of the convergence
of the hole-line expansion. The BHF approach has been extended to hyperonic matter by several authors
\cite{hans,isaac2011,bhf}. The interested reader is referred to these works for the specific 
details of the BHF calculation of hyperonic matter, and to Ref.\ \cite{bbg} for an extensive review of 
the BBG many-body theory.

 
\section{Quark matter nucleation in hadronic stars}
\label{sec:pe}
The conditions of phase equilibrium, in the case of a first-order phase transition 
\cite{footnote}, 
are given by the Gibbs' phase rule, 
which in the case of cold ($T = 0$) matter can be written as:  
\begin{equation} 
   P_H = P_Q \equiv P_0 \, ~~~~~~~~~~~~~~~ \mu_H(P_0)  = \mu_Q(P_0) \,
\label{eq:p0}
\end{equation}
where
\begin{equation}
\mu_H = \frac{\epsilon_H+P_H}{n_H} \, ~~~~~~~~~~  \mu_Q=\frac{\epsilon_Q+P_Q}{n_Q} 
\label{eq:mu}
\end{equation} 
are the Gibbs energies per baryon for the hadron (H) and quark (Q) phases, 
respectively, and the quantities $\epsilon_H (\epsilon_Q)$, $P_H (P_Q)$, and $n_H (n_Q)$ 
denote respectively the total ({\em i.e.,} including leptonic contributions) energy density, 
total pressure, and baryon number density of the two phases. 
Above the transition pressure $P_0$ the hadronic phase is metastable, and the stable quark 
phase will appear as a result of a nucleation process. 
Quantum fluctuations will form virtual drops of quark matter. The characteristic 
oscillation time $\nu_0^{-1}$ of these drops, in the potential energy barrier separating  the 
metastable hadronic phase and quark phase, is set by strong interactions, which are 
responsible of the deconfinement transition, thus $ \nu_0^{-1} \sim 10^{-23}$. 
This time is many orders of magnitude smaller than the weak interaction characteristic time 
($\tau_{weak} \sim 10^{-8}$ s), consequently quark flavor must be conserved forming a virtual 
drop of quark matter. We call $Q^*$-phase this deconfined quark matter, in which the flavor 
content is equal to that of the $\beta$-stable hadronic phase at the same pressure and 
temperature.  
Soon after a critical size drop of quark matter is formed, the weak interactions  
will have enough time to act, changing the quark flavor fraction of the deconfined 
droplet to lower its energy, and a droplet of $\beta$-stable quark matter is formed. 
(hereafter the Q-phase).

This first seed of quark matter will trigger the conversion \cite{oli87,hbp91,grb} of the 
pure hadronic star to a quark star. Thus, pure hadronic stars with values of the  
central pressure $P_c$ higher than $P_0$ and corresponding masses $M > M_{thr} \equiv M(P_0)$,  
are metastable to the decay (conversion) to quark stars \cite{b0,b1,b2,b3,b4,b5,b6,b7}.   
The mean lifetime of the metastable stellar configuration is related to the time needed to 
nucleate the first drop of quark matter in the stellar center, and it depends dramatically 
on the value of the stellar central pressure. 

As in Refs. \cite{b0,b1,b2}, we define as the {\it critical  mass} $M_{cr}$ of the hadronic star 
sequence, the value of the stellar gravitational mass for which the nucleation time of a 
$Q^*$-matter droplet is equal to one year: $M_{cr} \equiv M_{HS}(\tau = 1~{\rm yr})$.     
Pure hadronic stars with $M > M_{cr}$ are thus very unlikely to be observed. 
$M_{cr}$ plays the role of an effective maximum mass \cite{b2} for the hadronic branch 
of compact stars.  

In a cold and neutrino-free hadronic star the formation of the first drop of 
quark matter could take place solely via a quantum nucleation process.  
The basic quantity needed to calculate the nucleation time is the energy barrier separating 
the Q*-phase from the metastable hadronic phase. 
This energy barrier, which represents the difference in the free energy of the system with and 
without a Q*-matter droplet, can be written as \cite{iida98,b2}
\begin{equation} 
  U({\cal R}) = \frac{4}{3}\pi n_{Q^*}(\mu_{Q^*} - \mu_H){\cal R}^3 + 4\pi \sigma {\cal R}^2 \;,
\label{eq:potential}
\end{equation}
where ${\cal R}$ is the radius of the droplet (supposed to be spherical), and $\sigma$ is the 
surface tension for the surface separating the hadron from the Q*-phase. 
The energy barrier has a maximum at the critical radius 
${\cal R}_c = 2 \sigma /[n_{Q^*}(\mu_H - \mu_{Q^*})]$.    

The quantum nucleation time $\tau_q$ can be straightforwardly evaluated within a 
semi-classical approach \cite{iida98,b1,b2,b3} and it can be expressed as 
\begin{equation}
  \tau_q  = (\nu_0 p_0 N_c)^{-1} \ , 
\label{eq:time}
\end{equation} 
where $p_0$ is  the probability of tunneling the energy barrier $U({\cal R})$ in its ground state, 
$\nu_0$ is the oscillation frequency of a virtual drop of the Q*-phase in the potential well, and  
$N_c \sim 10^{48}$ is the number of nucleation centers expected in the innermost part 
($r \leq R_{nuc} \sim100$ m) of the hadronic star, where the pressure and temperature 
(in finite T case) can be considered constant and equal to their central values.

\section{Results and discussion}
\label{sec:results}

We will now discuss the results obtained with the two microscospic hadronic EoS considered and, 
in particular, we will comment whether the possible discussed scenarios are compatible with 
the recent measurement \cite{Demorest10} of the mass of the pulsar PSR J1614-2230  with 
a mass M = (1.97 $\pm$ 0.04) $M_\odot$. 
 
\begin{figure}[b]
 \centering
 \vspace{0cm}
 \begin{tabular}{cc}
 \includegraphics[width=1.\linewidth,angle=0,clip]{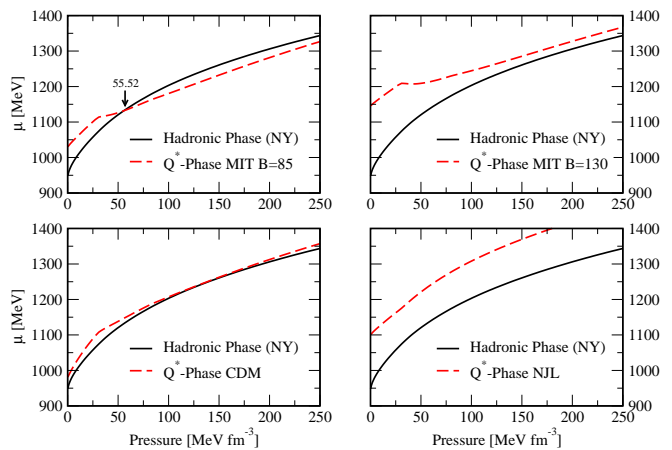}
 \end{tabular}
\caption{(Color online) Gibbs energy per baryon at zero temperature as a function of 
 the pressure for the hadronic phase (solid lines) and the $Q^{*}$-phase (dashed lines). 
 Model $1$ has been used to describe the hadronic phase. 
 Results for the MIT bag model with $m_u = m_d = 0, m_s = 150$~MeV and two 
 different values of the bag constant $B$, are presented in the
 top panels whereas those for the Chromo Dielectric (CDM) and the NJL models are 
 shown in the bottom ones. The arrow and the corresponding number indicates the 
 value (in MeV/fm$^3$) of the transition pressure $P_0$.}  
 \label{fig1}
 \end{figure}

In Fig. \ref{fig1} we plot the Gibbs energy per baryon as a function of pressure using the 
microscopic approach of Ref.\ \cite{hans} (Model $1$) for the hadronic phase and one of 
the following models for the $Q^*$-phase: 
MIT bag (top panels), CDM (left bottom panel) and NJL (right bottom panel) model.   
It is interesting to note that the formation of the $Q^*$-phase is possible only in the case 
of the MIT bag model EoS with a low value of the bag constant ($B=85$ MeV fm$^{-3}$).  
In all the other cases considered in Fig. \ref{fig1}, the curve for Gibbs energy per baryon 
for the $Q^*$-phase  never crosses the one for the hadronic phase, consequently, the hadronic 
phase will always remain stable with respect to the formation of $Q^*$-phase droplets. 
For these three QM models, this result implies that the pure hadronic stars 
(hyperonic stars) described by Model $1$ are stable up to their maximum mass configuration 
$M^{HS}_{max} = 1.37 M_\odot$.    

By numerical integration of the Tolman--Oppenheirmer--Volkov equations \cite{shapiro83}, 
we have calculated the structural properties for pure hadronic and quark star sequences.    
The main results, in the case of Model $1$ EoS for the hadronic phase, are summarized in 
Table \ref{I}, where we report 
the maximum gravitational mass $M^{HS}_{max}$ (third column), 
the gravitational threshold mass $M_{thr} \equiv M(P_0)$ for metastable configurations (fourth column), 
and the gravitational (baryonic) critical mass $M_{cr}$ ($M^b_{cr}$) (fifth (sixth) column) 
for the pure hadronic star sequence. 
$M_{fin}$ (seventh column) is the gravitational mass of the hybrid star formed by the stellar 
conversion process of the HS with $M = M_{cr}$ and assuming  baryon number conservation in 
the process \cite{grb} ({\it i.e.} assuming $M^b_{fin} = M^b_{cr}$). 
Finally  $E_{conv}$ is the total energy liberated in the stellar conversion. 
It is interesting to note that hybrid star configurations can be obtained with the MIT bag 
model with $B=85$~MeV fm$^{-3}$, and with the CDM.  In the latter case, however, the 
transitory non-$\beta$-stable $Q^*$-phase is not energetically achievable (see left bottom  
panel Fig. \ref{fig1}). Thus in this case the hadronic star sequence is stable up to the 
maximum mass configuration (thus we have no entries in Tab. \ref{I} for the quantities 
$M_{thr}$, $M_{cr}$, $M^b_{cr}$, $M_{fin}$ and $E_{conv}$). 
In the case of the MIT bag model with $B=85$~MeV fm$^{-3}$ the $Q^*$-matter nucleation 
is possible and  one has $M_{cr} = 1.272~M_\odot$. The conversion of this star will produce 
an hybrid star with $M_{fin} = 1.233~M_\odot$. If this object is a member of a binary stellar 
system, eventual accretion of matter from the companion will allow it to reach a maximum mass 
of 1.544 $M_\odot$. In this case the  pulsar PSR J1614-2230 will neither be an hyperonic star 
nor an hybrid star. 

We have also artificialy turned off the hyperonic degrees of freedom in Model 1 EoS, and 
considered pure nucleonic stars. In this case we have $M^{HS}_{max} = 2.27~M_\odot$, and 
a critical mass $M_{cr} \sim 2.2~M_\odot$ for all the quark matter EoS considered in Tab. \ref{I}. 
In all cases, however, the critical mass configuration will collapse to a black hole (BH entry in 
seventh column in Tab. \ref{I}) and thus the hybrid star sequence can not be populated. 
In this case the  pulsar PSR J1614-2230 would be an hadronic star containing only 
nucleons and leptons.
%
%
%
%
\begin{table*} 
 \begin{center}
 \bigskip                           
 \begin{ruledtabular}
 \begin{tabular}{llccccccc}
 &  & $M^{HS}_{max}$ & $M_{thr}$ & $M_{cr}$ & $M^b_{cr}$ & $M_{fin}$ & $E_{conv}$ & $M_{max}^{HybS}$ \\
 \hline   
nucleons+hyperons  
&MIT ($B=85$)    &  1.37   & 1.227 & 1.272& 1.397& 1.233 & 71.26  & 1.544\\  

&  CDM              &  1.37   &     -    &    -    &   -     &    -     &    -     & 1.591\\

only nucleons 
&MIT ($B=85$)     &  2.27   & 2.193 & 2.226& 2.677& BH & -  & 1.544\\
&MIT ($B=130$)   &  2.27   & 2.242 & 2.254& 2.720& BH & -  & 1.471\\
&NJL                  &  2.27   &2.229  & 2.246& 2.708& BH & -  & 1.879\\
&CDM                 &  2.27   & 2.242 & 2.255 & 2.722&BH &  - &1.592\\ 
 \end{tabular}
 \end{ruledtabular}
 \end{center} 
 \caption{ 
Stellar properties for the hadronic EoS of  Ref.\ \cite{hans} (Model $1$) in the case nuclear matter 
(only nucleons) or hyperonic matter (nucleons+hyperons) and different models for the quark phase. 
$M^{HS}_{max}$ is the gravitational maximum mass of the pure hadronic star 
sequence, $M_{thr} \equiv M(P_0)$  is the gravitational threshold mass for metastable stellar 
configurations, $M_{cr}$ ($M^b_{cr}$) the gravitational (baryonic) critical mass, 
$M_{fin}$ is the mass of the hybrid star which is formed by the stellar conversion of the hadronic 
star with $M=M_{cr}$ and assuming  baryon number conservation in the process 
({\it i.e.}  $M^b_{fin} = M^b_{cr}$), finally  $E_{conv}$ is the total energy, in unit of $10^{51}$~erg, 
liberated in the stellar conversion. All stellar masses are expressed in units of the solar mass
$M_\odot = 1.9889 \times 10^{33}$~g. The bag pressure $B$ is given in MeV/fm$^3$} 
 \label{I}
 \end{table*}
%
\begin{figure}[b]
\vspace{0cm}
\includegraphics[width=1.\linewidth,angle=0,clip]{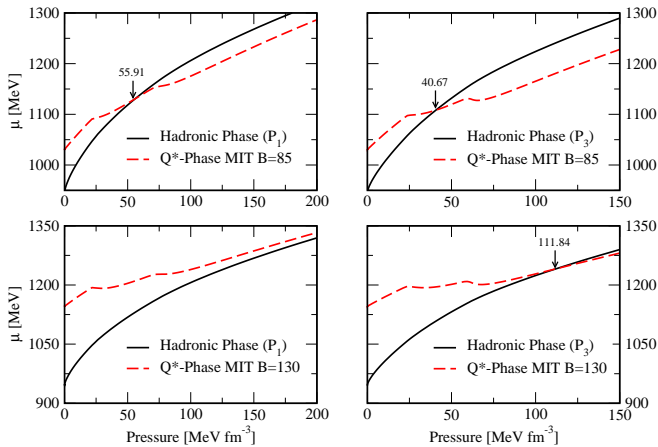}
\caption{(Color online) 
 Gibbs energy per baryon at zero temperature as a function of 
 the pressure for the hadronic phase (solid lines) and the $Q^{*}$-phase (dashed lines).  
The parametrizations $P_1$ (left panels) and $P_3$ (right panels) of the  Model $2$ 
introduced in \cite{isaac2011} have been used to describe the hadronic phase. 
Results for the MIT bag model with $B=85$ (top panels) and $130$ MeV/fm$^3$ (bottom panels) 
are presented. 
The arrow and the corresponding number indicates the value (in MeV/fm$^3$) of 
the transition pressure $P_0$.}\label{fig2}   
\end{figure} 
%
 \begin{table*} 
 \begin{center}
 \bigskip                           
 \begin{ruledtabular}
 \begin{tabular}{lcccccccc}
 &  & $M^{HS}_{max}$ & $M_{thr}$ & $M_{cr}$ & $M^b_{cr}$ & $M_{fin}$ & $E_{conv}$ & $M_{max}^{HybS}$ \\
 \hline       
  $P_1$ & MIT $B=85$  & 1.48  & 1.245 & 1.368 & 1.526 & 1.335 & 60.32 & 1.574 \\
  $P_2$ & MIT $B=85$  & 1.38  & 1.210 & 1.230 & 1.356 & 1.203 & 47.62 & 1.574 \\
  $P_3$ & MIT $B=85$  & 1.60  & 1.293 & 1.504 & 1.688 & 1.456 & 85.57 & 1.574  \\                      
  $P_3$ &     CDM        & 1.60  & 1.397 & 1.472 & 1.648 & 1.440 & 56.96 & 1.624  \\                
 \end{tabular}
 \end{ruledtabular}
 \end{center} 
\caption{Same as in Tab. I, but now for the hadronic EoS of  Ref.\ \cite{isaac2011} (Model $2$) and 
three different parametrizations of this model: 
$P_1$ ($K_\infty=236$ MeV and $x=1$), 
$P_2$ ($K_\infty=236$ MeV and $x=1/3$) and
$P_3$ ($K_\infty=285$ MeV and $x=1$).}
 \label{II}
 \end{table*}
%
%
We next discuss the results obtained with the hadronic EoS based on our recent work of 
Ref.\ \cite{isaac2011} (Model $2$) where, as we said, a microscopic Brueckner-Hartree-Fock 
approach of hyperonic matter based on the Argonne V18 NN and the NSC89 NY forces 
is supplemented with additional simple phenomenological density-dependent contact terms 
that mimic the effect of nucleonic and hyperonic three-body forces. 
In particular, we consider three different
parametrizations of this model corresponding to different values of the incompressibility 
coefficient, $K_\infty$, of symmetric nuclear matter at saturation, and the parameter $x$, which 
characterizes the strengh of the hyperonic three-body forces: $P_1$ ($K_\infty=236$ MeV and $x=1$), 
$P_2$ ($K_\infty=236$ MeV and $x=1/3$) and $P_3$ ($K_\infty=285$ MeV and $x=1$). 
The interested reader is referred to \cite{isaac2011}, and particularly to Tables 1 and 2 of 
this reference, for details.
The results for this model are shown in Table \ref{II} and Fig. \ref{fig2}. As in the previous case,
we summarize in Table \ref{II} the main stellar properties obtained with this model in 
combination with the MIT bag and the CDM. 
Note that in this case with the NJL model no transition occurs for any of the three parametrizations $P_1, P_2$ and $P_3$. 
In Fig. \ref{fig2} we plot the Gibbs energy per baryon as a function of pressure using: 
the MIT bag model for the $Q^*$-phase and an hadronic parametrization with the incompressibility 
$K_\infty=236$ MeV ($P_1$) and $K_\infty=285$ MeV ($P_3$) with $x=1$ in both cases. 
Results for the parametrization $P_2$, and the CDM and NJL models are not shown for conciseness.  
It is interesting to note that the parameter $x$, does not influence much the mass and 
radius of the hybrid star maximum mass configurations. 

Note also (see Table \ref{II}) that, similarly to Model $1$, for the parametrizations 
$P_1$ and $P_2$,  
the formation of the $Q^*$-phase is possible only in the case 
of the MIT bag model EoS with a low value of the bag constant ($B=85$ MeV fm$^{-3}$).  
Nevertheless, for the parametrization $P_3$, 
the stable hybrid sequence may be 
populated from the stellar conversion of 
the critical mass hadronic star    
if the quark phase is described either with the MIT bag model (with $B=85$ MeV/fm$^3$) 
or with the CDM model. As already said, no transition is found for the NJL model.   
We also note that in the case of the parametrization $P_3$ plus the MIT bag model with 
$B=130$ MeV/fm$^3$  
the $Q^*$-phase nucleation time at the center of the  maximum mass 
($M^{HS}_{max}$) hadronic star is much larger than the age of the Universe, and thus,  
it is extremely unlikely to populate the hybrid star branch in this case.  
In the most favorable scenario the possible largest star mass would be 1.624 $M_\odot$ (for CDM). 
This mass could occur if after, the conversion, the star accretes mass from an eventual companion 
star if the object is in a binary system.


\section{Summary and Conclusions}
\label{sec:conclusions}

Using the microscopic Brueckner--Hartree--Fock approach to describe the EoS
of dense hadronic matter we have studied the possibility of occurrence of a 
deconfinement phase transition into quark matter in neutron star cores. Quark matter 
has been described with three different models, namely, the MIT bag, the CDM and 
the NJL models. We have concluded that hyperonic hadronic stars will not suffer a 
deconfinement phase transition except if the quark EOS is 
obtained using  
 the  MIT bag model with  a value of the bag pressure $B=85$MeV/fm$^{3}$. In this case, 
however, it is not possible to get a star with a mass above 1.54 $M_\odot$.  On the 
other hand, we have found that if the hadronic matter has no hyperons then
deconfinement will occur only in very massive stars, with $M> 2.2 M_\odot$ and
the stars will decay into a black-hole. Within this microscopic approach to the
hadronic phase the  pulsar PSR J1614-2230 would be an hadronic star containing only 
nucleons and leptons.

We have also studied the possible effect of a hyperonic TBF using the model
proposed in \cite{isaac2011}. It was shown that for the hardest EOS with
$K_\infty=285$ MeV and $x=1$ an hybrid star could be formed. Only NJL and the
MIT with $B=130$ MeV/fm$^3$ did not predict a metastable star in this case. Within 
this scenario a maximum mass of 1.624 $M_\odot$ was predicted, very far from  the   
mass of the PSR J1614-2230. However, it was also shown that it was not so much
the hyperonic TBF strengh but more the incompressibility of the nucleonic
part of the EoS that defines the possible deconfinement transition.

  
\section*{Acknowledgments}
This work has been partially supported by the initiative QREN financed by the
 UE/FEDER throught the Programme COMPETE under the projects  
PTDC/FIS/113292/2009 and CERN/FP/116366/2010, the grant SFRH/BD/62353/2009, 
and by COMPSTAR, an ESF Research  Networking Programme.


 

\begin{thebibliography}{200}
\bibitem{footnote}{The deconfinement transition in the high density region relevant for neutron stars is 
assumed to be of first order.}
\end{thebibliography}
 \end{document}